\documentclass[12pt, a4paper,aps,amssymb,superscriptaddress,unsortedaddress]{article}
\usepackage{latexsym,amsmath,amssymb}
\usepackage{setspace}
\usepackage{graphicx}
\usepackage{psfrag}

\newcommand{\sw}{\stackrel{\star}{\wedge}}
\newcommand{\ep}{\epsilon}

\newcommand{\pd}{\partial}

\newcommand{\hphi}{\hat{\phi}}
\newcommand{\hbphi}{\hat{\bar{\phi}}}
\newcommand{\he}{\hat{e}}

\begin{document}

\topmargin -2pt


\headheight 0pt

\topskip 0mm \addtolength{\baselineskip}{0.20\baselineskip}

\vspace{5mm}

\begin{center}
{\Large \bf Noncommutative Solitonic Black Hole}
\vspace{10mm}

{\sc Ee Chang-Young}${}^{ \dag, }$\footnote{cylee@sejong.ac.kr},
{\sc Kyoungtae Kimm}${}^{  \dag,  }$\footnote{ktk@theory.sejong.ac.kr},
{\sc Daeho Lee}${}^{  \dag,  }$\footnote{dhlee@theory.sejong.ac.kr},
and {\sc Youngone Lee}${}^{  \ddag,  }$\footnote{youngone@daejin.ac.kr}
\\

\vspace{1mm}

 ${}^{\dag}${\it Department of Physics and Institute of Fundamental Physics,\\
Sejong University, Seoul 143-747, Korea}\\

${}^{\ddag}${\it Institute of Basic Sciences, Daejin University, Pocheon, Gyeonggi 487-711, Korea
}\\

\vspace{10mm}
{\bf ABSTRACT}
\end{center}


\noindent
\noindent

We investigate solitonic black hole solutions
in three dimensional
noncommutative spacetime.
We do this in gravity
with negative cosmological constant
coupled to a scalar field.
Noncommutativity is realized with the Moyal product
 which is expanded up to first order in
the noncommutativity parameter in two spatial directions.
With  numerical simulation
we study the effect of
 noncommutativity by increasing the
value of  the noncommutativity parameter
starting from commutative solutions.
We find that even a regular soliton solution in the commutative case
becomes a  black hole solution when the noncommutativity parameter reaches a certain value.

\vfill

\thispagestyle{empty}

\newpage


%

Many candidate theories for quantum gravity,
such as string theory and loop quantum gravity,
suggest that spacetime may not be commutative
at sufficiently high energy scales~\cite{Seiberg:1999vs,Freidel:2005me}.
Meanwhile,  black holes in the early universe
have been observed recently
\cite{Fan:2000bf,Willott:2007rm}.
Since the energy density of the early universe was very high,
it would be interesting
 to know
the effect of noncommutativity  on the formation of
a black hole.
Black holes in three dimensional spacetime have been extensively studied.
One of the reasons is that
gravity models in three dimensions are relatively easier to treat
than models in four spacetime dimensions.
The finding of the BTZ black hole solution~\cite{BTZ}
has raised a lot of interest in
the subject.

In~\cite{k3},
the global vortex solution  was studied
by considering gravity with negative cosmological constant
coupled to a complex scalar field
in three dimensional commutative spacetime.
There, the model Lagrangian
with a global $U(1)$ symmetry was given by
\begin{eqnarray}
\label{action_KKK}
S = \int d^3x\sqrt {-g}\left[-\frac{1}{16\pi G_N}(R+2\Lambda)\right.
+\left.\frac12 g^{\mu\nu}\partial_\mu\bar{\phi}\partial_\nu \phi
-\frac{\lambda}{4} (  \bar{\phi}\phi - v^2)^2
\right],
\end{eqnarray}
where $\phi(x)$ is a complex scalar field.
The obtained solution was a cylindrically symmetric global U(1) vortex solution
which  smoothly connects the false vacuum
 at the origin to the true vacuum  at spatial infinity.
Depending upon the ratio  of the cosmological constant
to the Plank scale,
the model supports a spacetime with a regular soliton
or a charged black hole.

In this paper, we investigate the effect of noncommutativity on the above model
and want to see whether  global vortex solutions with black hole configuration
are allowed in the noncommutative case.
 For this purpose we use the same Lagrangian as
in \eqref{action_KKK} except for the Moyal product
between the field variables.
 For computational purpose, here we use the triad and spin connection instead of the metric as in~\cite{ELL}.
 Since it is hard to obtain an analytic solution even in the commutative case~\cite{k3},
our approach to find the solution is basically numerical.
Our analysis is performed up to first order in
 the noncommutativity parameter.
\\
\indent
In this paper we work with the noncommutative polar coordinates
 $(\hat t,\hat r,\hat\varphi)$ defined by the following
commutation relation
\begin{equation}
\label{commrel:polar}
[\hat{\rho},\hat{\varphi}]=2i\theta,~~0~~\text{otherwise},
\end{equation}
where $\hat\rho\equiv \hat r^2$.
One reason for using the above commutation relation is that
 rotational symmetry is more apparent in the polar coordinates
than in the Cartesian coordinates.
Another quite important reason is that the above commutation relation is physically equivalent to
that of the canonical noncommutivity,
$[\hat{x},\hat{y}]=i\theta$,
up to first order in the nonocmmutativity parameter $\theta$~\cite{ELL}.
A would-be usual commutation relation for the noncommutative polar coordinates,
$[\hat{r},\hat{\varphi}]=i\theta$,
is not equivalent to the canonical noncommutivity,
$[\hat{x},\hat{y}]=i\theta$, even in the first order of $\theta$~\cite{ELL}.

By the Weyl-Moyal correspondence~\cite{Groe:46},
the physics in the noncommutative spacetime can be described
by the physics in the commutative spacetime with the Moyal product.
The Moyal product corresponding to the commutation relation \eqref{commrel:polar} is given by
\begin{eqnarray}
\label{Moyal_polar}
(f\star g)(\rho,\varphi)
= \left. e^{i\theta\left(
\frac{\partial}{\partial \rho}\frac{\partial}{\partial \varphi'}-\frac{\partial}{\partial \varphi}\frac{\partial}{\partial \rho'}
\right)}f(\rho,\varphi)g(\rho',\varphi')\right|_{(\rho,\varphi)=(\rho',\varphi')}.
\end{eqnarray}

Since the pure gravity action with negative cosmological constant can be written
as the Chern-Simons action~\cite{Achu:86},
the first two terms of the action~\eqref{action_KKK}
can be rewritten
in terms of the triad $e^a$ and the spin connection $\omega^a(a=0,1,2)$.
Then the noncommutative version of the action \eqref{action_KKK} can be written as
\begin{eqnarray}
\label{nc action}
\hat{S}= \frac{1}{8\pi G_N} \int \left(\hat{e}_a \sw \hat{R}^a+\frac{\Lambda}{6}\ep^{abc}\hat{e}_a\sw \hat{e}_b \sw \hat{e}_c \right)
+\int d^{3}x~ \hat{e}\star \hat{\cal{L}} [\hat{\phi}],
\end{eqnarray}
where $\hat{e}$ is the determinant of $\hat{e}_\mu^a$ and
$\hat{R}^a$ is the curvature 2-form, $\hat{R}^a  =  d\hat{\omega}^a + \frac{1}{2} \ep^{abc}~ \hat{\omega}_{b} \sw \hat{\omega}_{c}$,
and the Lagrangian ${\cal \hat{L}}[\hat{\phi}]$ for the scalar field is given by
\begin{eqnarray}
{\cal \hat{L}}[\hat{\phi}]
=-\frac{1}{4}(\partial_\mu \hat{\bar{\phi}}
 \star g^{\mu\nu} \star
 \partial_\nu \hat{\phi}
 +\partial_\mu \hat{{\phi}}
 \star g^{\mu\nu} \star
 \partial_\nu \hat{\bar{\phi}}
 )
-\frac{\lambda}{4}
(\hat{\bar{\phi}}\star \hat{\phi} - v^2)
\star
(\hat{\bar{\phi}}\star\hat{\phi}-v^2).
\end{eqnarray}
We define the noncommutative metric as
\begin{equation}
\label{def:ncmetric}
\hat{g}_{\mu\nu}
=\frac{1}{2}\eta_{ab}(
\hat{e}_{\mu}^a \star \hat{e}_{\nu}^b
+ \hat{e}_{\nu}^b \star \hat{e}_{\mu}^a)
\end{equation}
such that it is real and symmetric.
The equations of motion are as follows:
\begin{align}
\label{nc einstein}
&\frac{\ep^{\mu\nu\rho}}{8\pi G_N}
\left[ \hat{R}_a + \frac{\Lambda}{2}\ep_{abc}\hat{e}^b \sw \hat{e}^c
\right]_{\nu\rho} +
\frac{1}{6}\ep^{\mu\nu\rho}\ep_{abc}
\Big( \hat{e}^{b}_{\nu}\star \hat{e}^{c}_{\rho}\star \hat{\mathcal{L}}[\hat{\phi}]
+\he^{c}_\rho\star\hat{\mathcal{L}}[\hat{\phi}]\star\he^{b}_\nu
+\hat{\mathcal{L}}[\hat{\phi}]\star\he^{b}_\nu\star\he^{b}_\mu
\Big) \nonumber\nonumber \\
&+\frac{1}{8}
\Big[ \he^{\mu}_c\star\he^{\nu}_b\star\pd_{\nu}\hphi\star\he\star\pd_{\rho}\hbphi\star\he^{\rho}_a
+\he^{\mu}_c\star\pd_{\nu}\hphi\star\he\star\pd_{\rho}\hbphi\star\he^{\nu}_b\star\he^{\rho}_a
\cr
&+\he^{\mu}_b \star\pd_{\nu}\hphi\star\he\star\pd_{\rho}\hbphi\star\he^{\rho}_c\star\he^{\nu}_a
+\he^{\mu}_b \star\he^{\rho}_c\star\pd_{\nu}\hphi\star\he\star\pd_{\rho}\hbphi\star\he^{\nu}_a
 +(\hphi \leftrightarrow \hbphi)
\Big]=0, \\
\label{nc torsion}
& \hat{T}^a  \equiv  d\hat{e}^a+\frac{1}{2}\epsilon^{abc}(\hat{\omega}_b\sw\hat{e}_{c}+\hat{e}_b\sw\hat{\omega}_{c} )=0,
\\
\label{nc scalar}
&\pd_{\mu}(\hat{g}^{\mu\nu}\star \pd_{\nu}\hat{\phi}\star\hat{e}+\hat{e}\star\pd_{\nu}\hat{\phi}\star\hat{g}^{\mu\nu})
= \lambda\hat{\phi}\star
(\hat{\bar{\phi}}\star\hat{\phi}\star\hat{e}+\hat{e}\star\hat{\bar{\phi}}\star \hat{\phi}
 -2v^2\hat{e}).
\end{align}
In the commutative limit, $\theta \rightarrow 0$,
the equations (\ref{nc einstein})-(\ref{nc scalar})
reduce to the commutative ones~\cite{Carl:95,k3} as expected.
\\
\indent
In the commutative case,  a static and rotationally symmetric metric
can be put into the following form~\cite{k3}:
\begin{equation}
\label{metric:r}
ds^2=-e^{2A(r)}B(r)dt^2+\frac{dr^2}{B(r)}+r^2 d\varphi^2.
\end{equation}
The triad and spin connection corresponding to this line element are given by~\cite{Carl:95}
\begin{eqnarray}
&&e^0 = e^{A} \sqrt{B}~dt,~~~
e^1 = \frac{1}{\sqrt{B}} dr,~~~
e^2 = r d\varphi
\nonumber
\\
&&\omega^0= -\sqrt{B} d\varphi,~~~
\omega^1  = 0,~~~
\omega^2=
-\sqrt{B} \frac{d}{dr} \left( e^{A} \sqrt{B}  \right)dt.\nonumber
\end{eqnarray}
In terms of
$A$, $B$, and $\phi$, where $\phi\equiv |\phi|(r)e^{in\varphi}$, the commutative equations of motion reduced
 from (\ref{nc einstein})-(\ref{nc scalar}) are given by
\begin{align}
&A' = 8\pi G_N  r(|\phi|')^2,
\label{cmeqs}
 \cr
&B'= 2|\Lambda| r -8\pi G_N r \left[
B (|\phi|')^2 + \frac{n^2}{r^2} |\phi|^2 +\frac{\lambda}{2}(|\phi|^2-v^2)^2\right], \cr
&|\phi| '' +\left(A' + \frac{B'}{B} +\frac{1}{r}\right) |\phi|'
 = \frac{1}{B}\left(\frac{n^2}{r^2} + \lambda (|\phi|^2-v^2)\right)|\phi|,
\end{align}
and these are exactly the same equations that appeared in~\cite{k3}.


In line with the Weyl-Moyal correspondence and
taking a hint from the commutative metric (\ref{metric:r}),
  we now  make an ansatz
  for the noncommutative metric
 in the $(t,\rho,\varphi)$ coordinates
which satisfy the commutation relation (\ref{commrel:polar})
as follows:
%
\begin{equation}
\label{ncmetric:R}
d\hat{s}^2=-e^{2\hat{A}(\rho)}\hat{B}(\rho)dt^2+\frac{d\rho^2}{4\rho \hat{B}(\rho)}+\rho~d\varphi^2.
\end{equation}
A noncommutative triad
for  the above metric
compatible with the definition of the metric \eqref{def:ncmetric} can be chosen as
\begin{equation}
\label{nctriads}
\hat{e}^0=e^{\hat{A}(\rho)}\sqrt{\hat{B}(\rho)}dt,~\hat{e}^1=\frac{d\rho}{\sqrt{4\rho\hat{B}(\rho)}},~\hat{e}^2=\sqrt{\rho} d\varphi.
\end{equation}
With the above choice of the triad,
the noncommutative spin connection can be determined
from the noncommutative torsion free condition (\ref{nc torsion}):
\begin{eqnarray}
\label{ncspincns}
\hat{\omega}^0=-\sqrt{\hat{B}(\rho)}d\varphi,~\hat{\omega}^1=0,~
\hat{\omega}^2
=-\sqrt{4\rho\hat{B}(\rho)}
\frac{d}{d\rho}\left(e^{\hat{A}(\rho)}\sqrt{\hat{B}(\rho)}\right)dt.
\end{eqnarray}
\begin{figure}
\centering
    \includegraphics[width=0.55\textwidth]{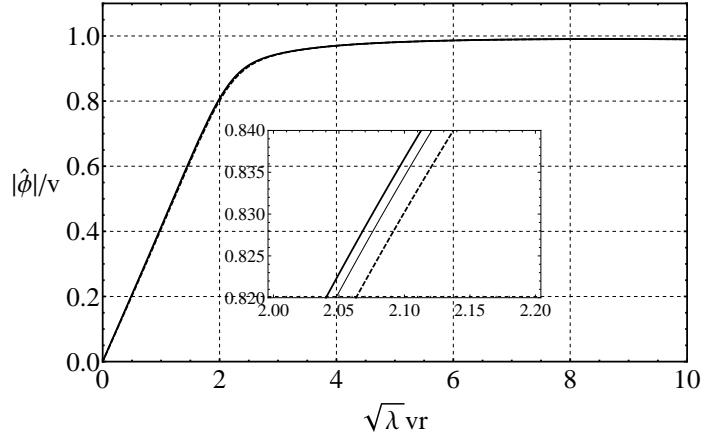}
  \caption{The noncommutative vortex solutions
  with $\Lambda_v=0.1$ and $ G_v =1.33$ are plotted
  for the scalar field $|\hat{\phi}|/v$.
   The dashed line is for the case of $\theta=0$. The thin and thick solid lines are for the cases of $\theta=0.1, 0.15$, respectively.}
  \label{figone}
\end{figure}

Now we expand the metric functions $\hat{A}, \hat{B},$ and
the scalar field $\hat{\phi}$ for static global vortices with vorticity $n$
 in terms of  $\theta$ up to first order  as follows.
\begin{eqnarray}
\label{ncexpand}
e^{2\hat{A}(\rho)}\hat{B}(\rho)&=&e^{2\tilde{A}(\rho)}\tilde{B}(\rho)+\theta~\tilde{F}(\rho)+\mathcal{O}(\theta^2),
\nonumber \\
\hat{B}(\rho)&=&\tilde{B}(\rho)+\theta~\tilde{G}(\rho)+\mathcal{O}(\theta^2),
\nonumber\\
\hat{\phi}(\rho,\varphi)&=&(\tilde{\phi}(\rho)+\theta~\tilde{\Phi}(\rho))e^{in\varphi}+\mathcal{O}(\theta^2).
\nonumber
\end{eqnarray}
With the following redefinitions of functions,
\begin{eqnarray}
&&  \tilde{A}(\rho)  \equiv A(r), ~\tilde{B}(\rho)  \equiv B(r),
~\tilde{\phi}(\rho)  \equiv |\phi|(r) ,  \nonumber \\
&& \tilde{F}(\rho) \equiv F(r), ~\tilde{G}(\rho)  \equiv G(r),
~\tilde{\Phi}(\rho) \equiv \Phi(r),\nonumber
\end{eqnarray}
and from (\ref{nc einstein})-(\ref{nc scalar}),
we get the original commutative equations \eqref{cmeqs} in the zeroth order of $\theta$,
and the following three equations for $ F(r),  G(r), \Phi(r) $
 in the first order of $\theta$:
\begin{align}
&a_1(r)F(r)
+a_2(r)F'(r)
+a_3(r)G(r)
+a_4(r)G'(r)
\cr
&+a_5(r)\Phi(r)
+a_6(r)\Phi'(r)
+a_7(r)\Phi''(r)
+a_8(r)=0,
\cr
&b_1(r)G(r)
+b_2(r)G'(r)
+b_3(r)\Phi(r)
+b_4(r)\Phi'(r)
+b_5(r)=0,
\label{nceq3}
\cr
&c_1(r)F(r)
+c_2(r)F'(r)
+ c_3(r)G(r)
+c_4(r)\Phi(r)
+c_5(r)\Phi'(r)
+c_6(r)=0.
\end{align}
All the coefficients in these equations
are functions of the known commutative solution, $|\phi|$, $A$ and $B$,
and are given in the appendix.

The numerical analysis was performed
up to first order in the noncommutativity parameter $\theta$
for the vorticity $n=1$.
The cosmological constant,
the Newton constant, and the radial coordinate are scaled to
$\Lambda_v=|\Lambda|/\lambda v^2$,  $G_v = 8\pi G_N v^2$, and $\sqrt{\lambda}vr$, respectively.
\begin{figure}[!b]
 \centering   \includegraphics[width=0.55\textwidth]{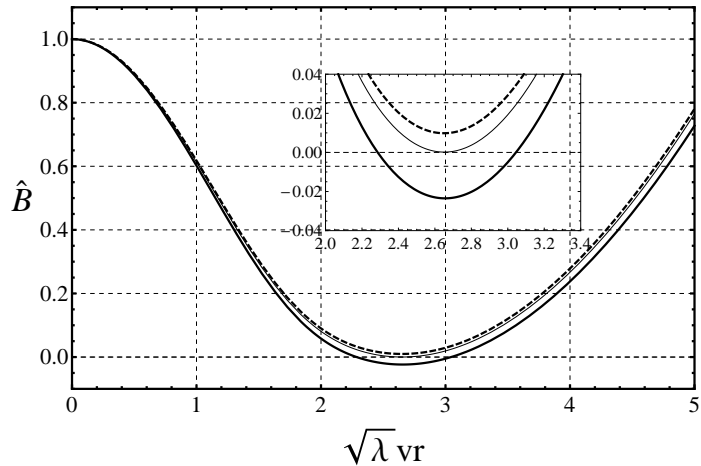}
  \caption{The solutions for noncommutative metric are plotted
  for $\hat{B}$ at values $\Lambda_v=0.088$ and $G_v =1.33$.
  The dashed line is for the case of $\theta=0$.
  The thin and thick solid lines are for the cases of $\theta=0.057, 0.2$, respectively.
   We note that there is a fine split at $r=0$.
  }
  \label{F_ncBr}
\end{figure}
In order to solve (\ref{nceq3}) we impose the Dirichlet condition at the origin
for $F$ and $G$ which is compatible with the
boundary condition for $\Phi(r)$
\begin{equation}
\label{ncbc}
\Phi(0)=\Phi(\infty)=0.\nonumber
\end{equation}
\begin{figure}[t]
  \centering
  \includegraphics[width=0.55\textwidth]{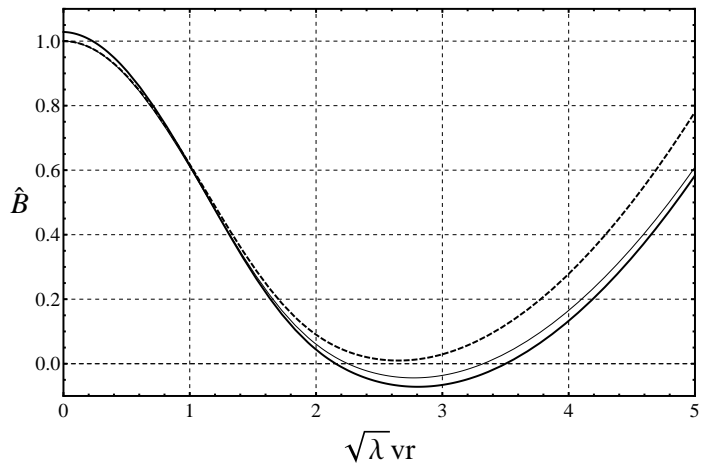}
  \caption{The noncommutative black hole solutions
  with $G_v =1.33$ are plotted
  for  $\hat{B}$.
  The dashed and thin solid lines are for the cases
  of $\Lambda_v=0.087, 0.08$, respectively, at  $\theta=0$.
  The thick solid line is for the case of $\Lambda_v=0.08$
  at $\theta=0.2$.}
   \label{ncBr_bh}
  \end{figure}
\begin{figure}[t]
  \centering
  \includegraphics[width=0.55\textwidth]{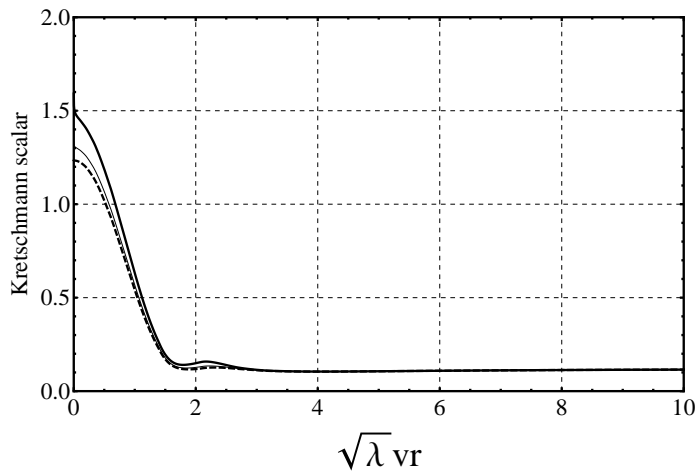}
  \caption{The plot of the noncommutative Kretschmann scalar with $\Lambda_v=0.1$
  and $G_v =1.33$.
   The dashed line is for the case of $\theta=0$.
   The thin and thick solid lines are
    for the cases of $\theta=0.057, 0.2$, respectively.
  }
   \label{ncKretschmann}
  \end{figure}
The effects of turning on $\theta$ on the global vortex solutions is shown in Fig.~\ref{figone}. This shows that the scalar field concentrates inwards as $\theta$ grows.
The gradient of the scalar field around $r=0$ becomes steeper.
In Fig.~\ref{F_ncBr}, the metric function $\hat{B}$
corresponding to the global vortices in Fig.~\ref{figone} is drawn.
Note that the commutative solution ($\theta=0$) has no horizon.
On the other hand, when the value of the noncommutativity parameter $\theta$ reaches $0.057$,
the solution becomes an extremal black hole solution.
When the value of $\theta$ becomes larger than this value, for instance  $\theta=0.2$,
  it becomes a nonextremal black hole solution.
When the
starting commutative solution
is a nonextremal black hole solution with $\Lambda_v=0.08$, the corresponding
noncommutative solution for  $\theta=0.2$ is shown in Fig.~\ref{ncBr_bh}.
The effect of turning on $\theta$ for  already-black holes
in commutative spacetime is shown.
As the noncommutativity parameter $\theta$ gets bigger,
 the area of the outer horizon increases while the separation between the inner
 and outer horizons grows.
 This effect is what we would expect when the mass of a nonextremal black hole
increases in the commutative case.

In order
to see whether the singularities of the metric at zeros
are  coordinate artifacts or true physical singularities, the Kretschmann scalars
$\hat{R}_{\mu\nu\rho\sigma}\hat{R}^{\mu\nu\rho\sigma}$ for the solutions having one or two zeros in Fig.~\ref{F_ncBr}
are plotted in Fig.~\ref{ncKretschmann}.
This shows that the Kretschmann invariant for the regions around these zeros
behaves regularly.
 From this result we can say that the zeros
 are not genuine singularities,
rather they correspond to the horizons of black holes.

\begin{figure}[t]
 \centering
 \includegraphics[width=0.6\textwidth]{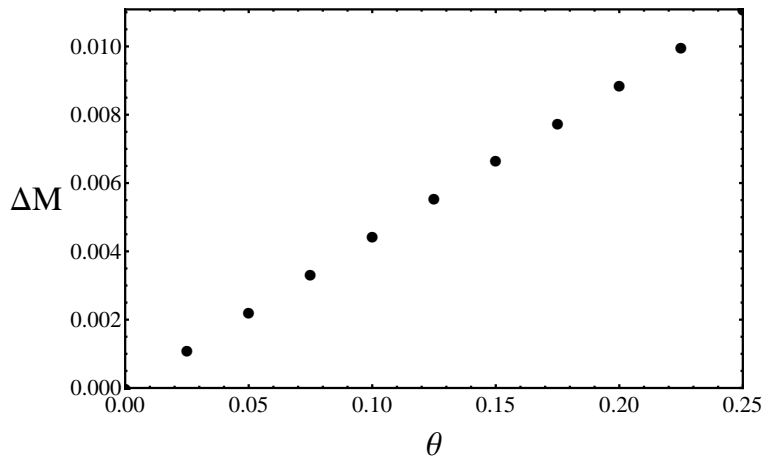}
 \caption{The plot of the gravitational mass increase vs $\theta$ with $\Lambda_v=0.1$ and $G_v =1.33$.}
 \label{F_ncADM}
\end{figure}

The increase in gravitational mass
for different values of $\theta$
by using the Hamiltonian formalism of~\cite{Hawking:1995fd} is plotted in Fig.~\ref{F_ncADM}.
The increase in gravitational mass for a given $\theta$
is defined by
\begin{equation}
\Delta M \equiv  H(\theta \ne 0) - H(\theta = 0), \nonumber
\end{equation}
where $H$ denotes the Hamlitonian.
The result shows that the gravitational mass increases
as the noncomutativity parameter $\theta$ increases.
This linearity shown in  Fig.~\ref{F_ncADM} is due to our analysis which was performed up to first order in $\theta$.
The gravitational mass defined here depends only on
the asymptotic geometrical quantities at spatial infinity.
The metric is fully determined by the scalar field and its derivatives
which are constant at spatial infinity.
Since the coefficient of $\theta$ is constant,
it implies the linear dependence of mass on $\theta$.

Thus our result sums up as follows.
The `inward' behaviour of the global vortex in  Fig.~\ref{figone}
results in higher peaks of the Kretschmann scalar near $r=0$
as in Fig.~\ref{ncKretschmann}.
These peaks correspond to gravitational  energy concentrations.
We may interpret this as the noncommutativity of space
makes the scalar soliton have higher concentration near the center than in the commutative case.
This concentrated scalar soliton behaves as
   the concentration of gravitational energy to
such an extent that a  black hole can form.

In the space of solutions,
a solution with no singularity changes into a black hole solution
as $\theta$ increases.
In other words, as one increases the value of the noncommutativity parameter,
the spacetime starts to allow a black hole solution at a certain value of the noncommutativity parameter.
This is not what we expected before.
A `phase change' in the solution space happens
as the noncommutativity parameter changes.

Comparing with the result obtained in~\cite{ELL}
for noncommutative BTZ black hole,
the role of the scalar field in the present work on forming a black hole is similar to that of the magnetic flux $B$ there.
%
The shift of the locations of the horizons of the noncommutative black holes in that paper
depends upon both $\theta$ and the magnetic flux $B$ at the origin.
The  shift of horizons obtained in this work
depends on both $\theta$ and the scalar field.
Thus, one may infer that
 the global vortex around the origin
takes over the role of the magnetic flux in~\cite{ELL}.
\\

\section*{Acknowledgments}

This work was supported by the National Research Foundation (NRF) of
Korea grants funded by the Korean government (MEST)
[R01-2008-000-21026-0 and NRF-2009-0075129 (E.\ C.-Y., K.\ K. and D.\ L.), and The Korea Research Foundation Grant funded by the Korea Government(MOEHRD), KRF-2008-314-C00063(Y.L.)

\section*{Appendix}

The coefficients appearing in the equations \eqref{nceq3} for $F(r), G(r), \Phi(r)$  are given
as follows:
\begin{align}
a_1(r) &=-r\Big[ (n^2-\lambda v^2 r^2)\phi+\lambda r^2 \phi^3-rB
( (1-rA')\phi'+r\phi'')\Big],\cr
a_2(r) &= r^3B\phi', \cr
a_3(r) &=r e^{2A}\Big[
(n^2-\lambda v^2 r^2)\phi+\lambda r^2 \phi^3+rB( (1+rA')\phi'+r\phi'')
\Big], \cr
a_4(r) &=e^{2A}r^3 B\phi', \cr
a_5(r) &= -2e^{2A}rB(n^2-\lambda v^2 r^2+3\lambda r^2\phi^2), \cr
a_6(r) &= 2e^{2A}r^2 B (B+rBA'+rB'), \cr
a_7(r) &= 2e^{2A}r^3B^2, \cr
a_8(r) &=nBe^{2A}\Big[
(n^2+2\lambda v^2 r^2)\phi A'-2\lambda r^2 \phi^3 A'
-r^2( A'B'\phi'+B( A'^2\phi'+\phi'A''+A'\phi''))\Big], \cr
b_1(r) &=\frac{1}{2}\Big[
r(B'-2\Lambda r )+4\pi G_N (
2(n^2-\lambda v^2r^2)\phi^2+\lambda r^2\phi^4+r^2(\lambda v^4-2B\phi'^2))\Big], \cr
b_2(r) &=-rB, \cr
b_3(r) &=-16\pi G_NB\phi(n^2-\lambda v^2r^2+\lambda r^2\phi^2), \cr
b_4(r) &=-16\pi G_N r^2B^2\phi', \cr
b_5(r) &= 8\pi G_N n\lambda rB\phi (\phi^2-v^2)\phi', \cr
c_1(r) &=\frac{1}{2}\Big[
r(2\Lambda r+2BA'+B')-4\pi G_N(
2(n^2-\lambda v^2r^2)\phi^2+\lambda r^2\phi^4+r^2(\lambda v^4-2B\phi'^2))\Big], \cr
c_2(r) &=-rB, \cr
c_3(r) &=-re^{2A}( B'+2B(A'-4\pi G_N r\phi'^2)), \cr
c_4(r) &= -16\pi G_N e^{2A}B\phi(n^2-\lambda v^2r^2+\lambda r^2\phi^2), \cr
c_5(r) &=16\pi G_N e^{2A}r^2B^2\phi', \cr
c_6(r) &= 8\pi G_N e^{2A}n\lambda rB\phi(\phi^2-v^2)\phi'. \nonumber
\end{align}

\end{document}